# Ultra-broadband photodetectors based on epitaxial graphene quantum dots

Abdel El Fatimy[†*], Anindya Nath[‡], Byoung Don Kong[‡], Anthony K. Boyd[‡], Rachael L. Myers-Ward[‡], Kevin Daniels[‡], M. Mehdi Jadidi[§], Thomas E. Murphy[§], D. Kurt Gaskill[‡] and Paola Barbara[†*]

[†]Department of Physics, Georgetown University, Washington, DC 20057, USA.

[‡]US Naval Research Laboratory, Washington, DC 20375, USA.

[§]Institute for Research in Electronics and Applied Physics, University of Maryland, College Park, MD 20742 USA.

(Dated: August 30, 2017)

## Abstract

Graphene is an ideal material for hot-electron bolometers, due to its low heat capacity and weak electron-phonon coupling. Nanostructuring graphene with quantum dot constrictions yields detectors with extraordinarily high intrinsic responsivity, higher than $1\times10^9$ V W$^{-1}$ at 3K. The sensing mechanism is bolometric in nature: the quantum confinement gap causes a strong dependence of the electrical resistance on the electron temperature. Here we show that this quantum confinement gap does not impose a limitation on the photon energy for light detection and these quantum dot bolometers work in a very broad spectral range, from terahertz, through telecom to ultraviolet radiation, with responsivity independent of wavelength. We also measure the power dependence of the response. Although the responsivity decreases with increasing power, it stays higher than $1\times10^8$ V W$^{-1}$ in a wide range of absorbed power, from 1 pW to 0.4 nW.

---

[*] e-mail: a.elfatimy@gmail.com, Paola.Barbara@georgetown.edu.

Graphene is a broadband light absorber because it is a gapless material [1,2]. At low frequencies, up to the terahertz range, light absorption mainly occurs via intraband transition and it is determined by the graphene Drude conductivity [3,4]. At frequencies above the infrared range, interband optical transitions dominate, and light absorption reduces to a constant value, about 2.3% per graphene layer [4,5]. In all cases, electrons thermalize via electron-electron interactions within a time scale of tens of femtoseconds [6-8] and via emission of optical phonons within a few hundreds of femtoseconds [9,10]. After thermalization, the effective electron temperature $T_e$ can be higher than the temperature of the graphene lattice, of the substrate and of the metal contacts attached to the graphene, due to the small electronic heat capacity and the ineffective cooling from collisions with acoustic phonons [11,12].

There have been several demonstrations of graphene detectors based on hot electrons. Photothermal effect detectors use an asymmetric device architecture (PN junctions [13,14] or contacts made of different materials [15]) to produce a net current of hot electrons. For symmetric graphene devices, the increase in electron temperature can be measured either via Johnson noise thermometry [16,17] or by using the variation of the graphene resistance as a function of temperature [18,19]. In this latter case, the detector performance directly depends on how strongly the graphene resistance varies as a function of temperature. One important figure of merit is the bolometer responsivity, defined as the change of voltage $\Delta V_{DC}$ across the device caused by the incident light divided by the absorbed power, $r = \Delta V_{DC}/\Delta P = I_{DC}(\Delta R/\Delta P) = (I_{DC}/G_{TH})(\Delta R/\Delta T)$, where $G_{TH}$ is the thermal conductance, $\Delta R$ is the change in resistance caused by a temperature increase $\Delta T$ and $\Delta V_{DC}$ is measured at a constant current $I_{DC}$.

We recently showed that nanostructured quantum dot constrictions in epitaxial graphene grown on SiC yield THz detectors with extraordinarily high responsivity[20]. Detection of 2-mm-wavelength (150 GHz) radiation from quantum dots with different dot diameters revealed that the quantum confinement gap in the dot causes a strong temperature dependence of the graphene electrical resistance and therefore a high responsivity. The quantum confinement gap is a combination of the charging energy of the dot and the electronic level spacing, and it produces a potential barrier to the current flowing through the graphene [21-24]. The barrier heights extracted from fits of the temperature dependence of the resistance to a thermal activation behavior ranged from 0.5 meV to 4 meV for dot diameters varying from 200 to 30 nm. In all cases, the barrier height was larger than (or, for the largest dots, comparable to) the photon energy at 150 GHz (about 0.6 meV), leaving it unclear whether the quantum dot bolometers would operate at frequencies with photon energy higher than the barrier height and whether the performance would be frequency dependent. In this work, we study light detection from these graphene quantum dots up to photon energies a thousand times larger than the thermal activation barrier and show that the responsivity is completely independent of frequency. This can be explained by considering both charge carrier dynamics in graphene and quantum transport through the quantum dot. We also characterize the response as a function of absorbed power and find a sublinear dependence of the photovoltage on absorbed power, leading to a decrease of responsivity with absorbed power. The quantum dot devices nonetheless show responsivity higher than $1\times10^8$ V W$^{-1}$, at least three orders of magnitudes higher than the highest responsivity reported for other types of graphene bolometers [18,19], over a wide range of absorbed power, from 1pW to 0.4 nW.

The quantum dots were fabricated using e-beam lithography on epitaxial graphene grown on SiC and following the fabrication procedure described in Ref. [20,25]. The samples were mounted on a gold plate and placed in a cryostat behind a picarin vacuum window. Filters were inserted to control the optical input power and restrict the wavelength to eliminate thermal blackbody radiation.

Figure 1(a) shows the temperature dependence of the electrical resistance R(T) of a 100-nm diameter quantum dot (red curve). In our previous work [20] we characterized the performance of the graphene quantum dot bolometers as a function of temperature and dot diameter and showed that the smallest dots (with a diameter smaller than 100 nm) yield the best performance (responsivities higher than $1\times10^{10}$ V W$^{-1}$) because they have the largest quantum confinement gap and the strongest variation of resistance with temperature. In this work we focus on the study of the bolometer performance as a function of radiation wavelength and absorbed power, using dots of intermediate size, with diameters ranging from 100 nm to 200 nm. All the measurements presented here are performed at the base lattice temperature $T_0$= 3 K.

Figure 1(b) shows the photoresponse of a 200 nm dot. The black curve is the current-voltage (IV) characteristic of the dot with radiation off. The curve is nonlinear due to Joule heating: when the bias and the Joule power increase, the resistance decreases, as can be seen in the inset in Fig. 2(a). Under illumination, the resistance decreases further. The red, green and purple curves in Fig. 1(b) show the IV characteristics when the sample is irradiated with light at three different wavelengths ranging from millimeter-wave to ultraviolet (2 mm, 1543 nm and 365 nm, respectively). The sample clearly shows a response in this wide range of wavelengths. We analyzed the response by measuring the power absorbed by our devices from the IV curves. Figure 2 (a) shows the response of a device as a function of power for illumination at 2-mm wavelength (0.15 THz). For every value of incident power, we measure the power absorbed by the bolometer with the same method that we used in Ref. [20]. We first measure the differential resistance at zero bias for the current-voltage characteristic with radiation ON and then find the point in the current-voltage characteristic with radiation OFF that exhibits the same differential resistance. We use the Joule power dissipated in the bolometer at that point, $P = I_{DC}V_{DC}$, as a measurement of the radiation power absorbed when the light is ON. We repeat the same procedure for every IV curve that we obtain when varying the power of the incident light.

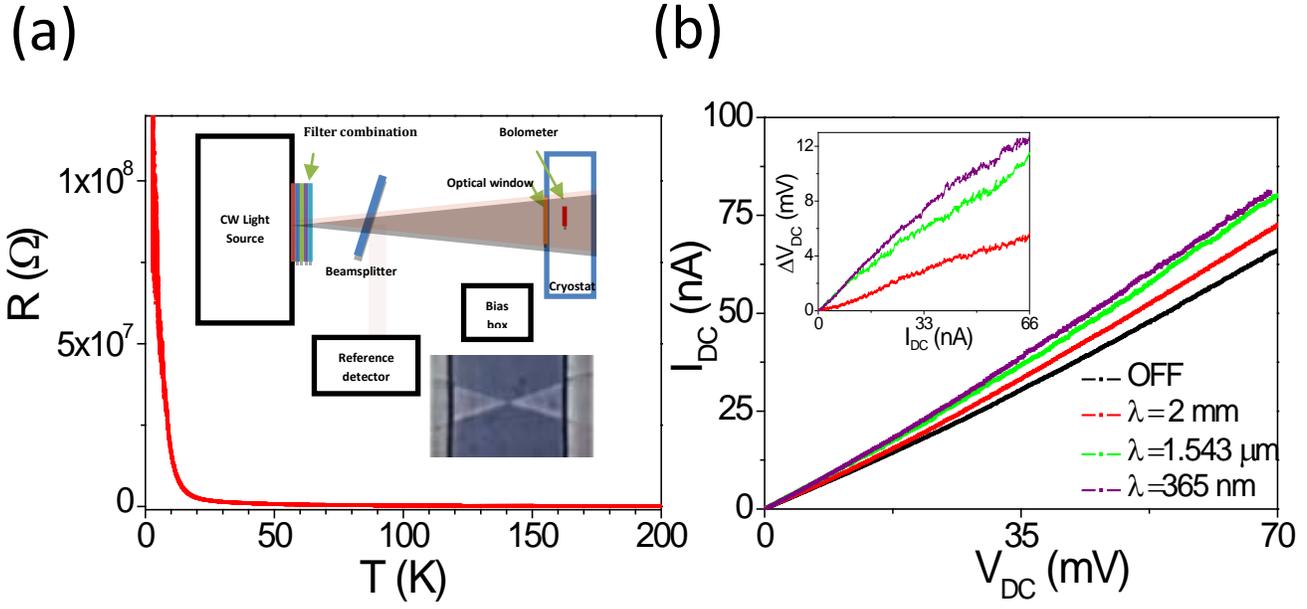

FIG. 1. (a) Resistance as a function of temperature for a quantum dot with a diameter of 100 nm. Inset: Optical image of a typical quantum dot and schematic of the experimental setup. (b) Current-Voltage characteristic of a 200-nm dot without radiation (OFF, black) and with radiation at 2 mm (red), 1.543 μm (green) and 365 nm (purple) wavelength, having absorbed power of 0.4, 1.0 and 1.4 nW, respectively. Inset: the response $\Delta V_{DC}$ as a function of the current $I_{DC}$ at the same wavelengths.

The photovoltage $\Delta V_{DC}$ at a fixed current, $I_{DC} = 1.9$ nA, for different values of the absorbed power is shown in Fig. 2(b) (black squares). The sublinear dependence of $\Delta V_{DC}$ (P) is expected because the variation of resistance vs. temperature is highest at low temperature (see Fig. 1(a)) and becomes weaker when the electron temperature increases due to an increase in the absorbed power. Since $\Delta V_{DC}$ is sublinear as a function of absorbed power, the responsivity decreases with increasing radiation power, but it is still very high for a large range of absorbed power, as shown in the inset in Fig. 2(b), where we include data from six devices, including two small-diameter devices with 30-nm dots.

The responsivity is larger for dots of smaller diameter. Figure 3 shows the power dependence of the responsivity for a 100-nm diameter dot. Measurements at the lowest values of absorbed power in Figure 3 (a) show that absorbed power below one picowatt can be detected at all the measured wavelengths by measuring changes in $V_{DC}$. Higher sensitivity can be achieved with temporally modulated illumination, using lock-in detection.

In all the measurements above, we have reported the absorbed power. Although this gives the ultimate performance of the devices, it is most useful to characterize their performance in terms of incident power. The optical coupling efficiency (the ratio between absorbed power and incident power), varies with radiation wavelength. For our long-wavelength source (2 mm, where the absorption is dominated by the graphene Drude conductivity), it can be optimized by designing antennas that are either broadband or tuned at specific wavelengths. For wavelengths of 1500 nm or shorter, the coupling efficiency is limited by the optical absorption of a single graphene layer, 2.3%. Our measured ratio of absorbed to incident power at 1543 nm is a bit higher, 2.7 %, (see Methods) and it is possibly enhanced by the reflection of radiation from the gold plate under the bottom surface of the SiC substrate.

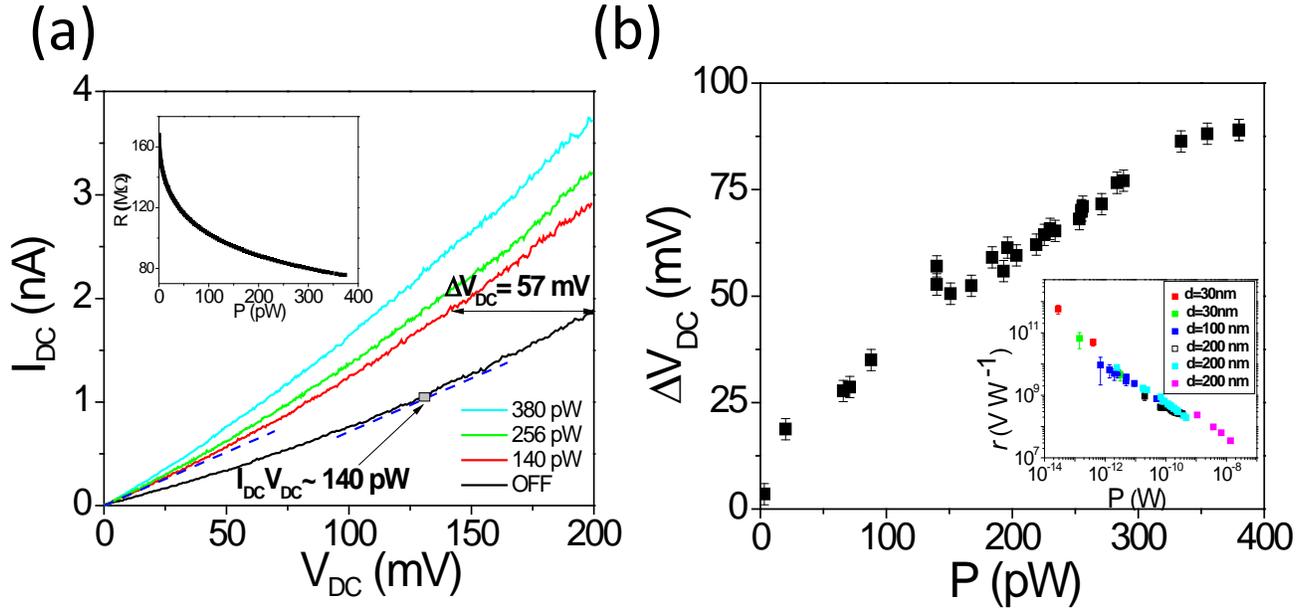

FIG. 2. (a) Current-voltage characteristic of a 200-nm dot without radiation (black) and with 2-mm wavelength radiation (colors) at different values of absorbed power. The dashed lines indicate points in the curves with the same differential resistance. Inset: Differential resistance of the curve with radiation OFF at different values of Joule power. (b) Photovoltage measured from the device in 2(a) at $I_{DC}$ = 1.9 nA as a function of absorbed power under illumination with a 2-mm wavelength source (black squares). The inset shows the responsivity vs. absorbed power for the same device (black squares) and five other devices. The legend indicates the diameter of the quantum dot for each device.

We use the measured responsivity at different wavelengths to calculate the total electrical NEP for the bolometers, including contributions from Johnson noise, shot noise and thermal fluctuations[26], ($NEP^2 = NEP^2_{JN} + NEP^2_{SN} + NEP^2_{TF} = (4k_BTR)/r^2 + (2eI_{DC})R^2/r^2 + 4k_BT^2G_{TH}$, where $G_{TH}$ is extracted from the data using the IV and R(T) curves [20]). Figure 3 indicates that the responsivity and the NEP are independent of radiation wavelength for photon energies in a very wide range of the radiation spectrum, including photon energies that are orders of magnitude larger than the activation energy of the dot.

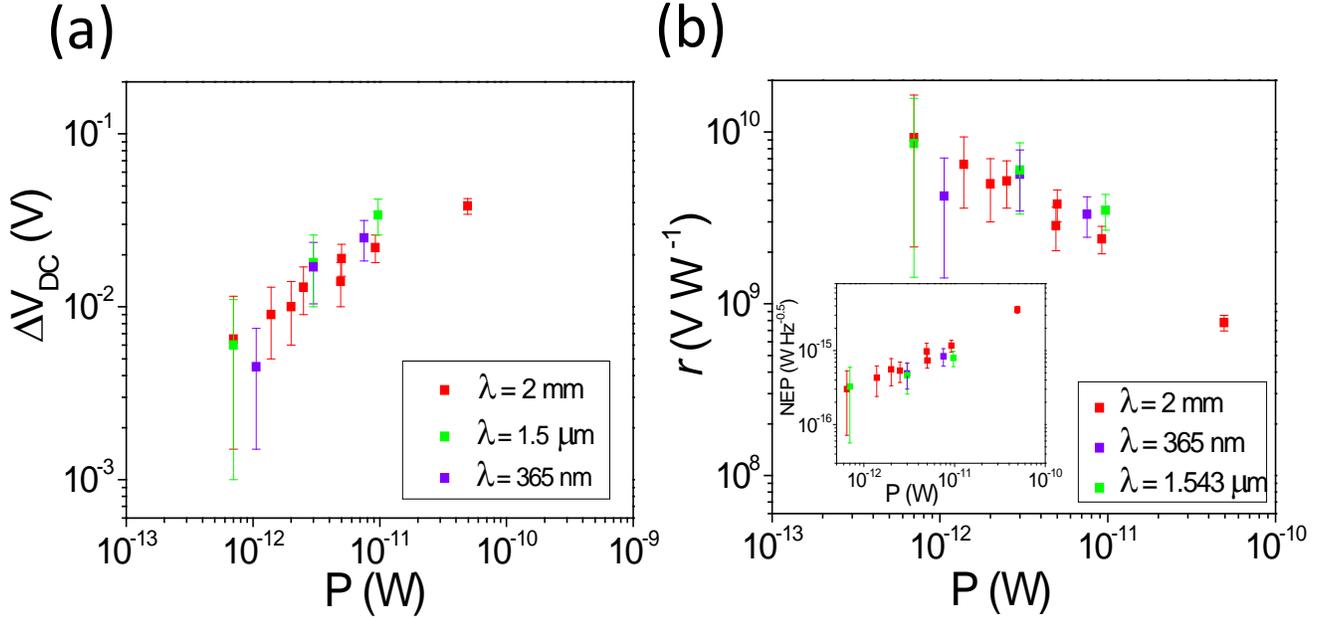

FIG. 3. (a) Low power dependence of the response from a 100-nm dot under illumination at different wavelengths. (b) Responsivity as a function of absorbed power at different wavelengths. Inset: Calculated electrical noise equivalent power (NEP) as a function of absorbed power at various wavelengths.

These behaviors can be explained by considering charge carrier dynamics after light absorption. As mentioned earlier, the timescale for electrons to equilibrate at an effective electron temperature via electron-electron collisions and optical phonon emission is on the order of 10 to 100 femtoseconds. This timescale is extremely fast compared to the charging time of the dot. We can estimate the capacitance of the dot by considering that the charging energy must be larger than the activation energy of the dot. This is because the activation energy depends on the alignment of the Fermi energy of the source and drain electrodes within the quantum confinement gap (it can be tuned to zero by doping the dot and aligning the Fermi energy to the top of the quantum confinement gap). The capacitance corresponding to a charging energy of 10 meV is about 10 aF, giving an RC time of 1 ns for a dot resistance of 100 MΩ. Since electrons equilibrate with a much faster timescale, the photon energy, and the specific wavelength do not play any role in the quantum transport of charges through the dot. As a result, we find that the absorbed power (regardless of wavelength) and the corresponding electron temperature are the only factors that determine the current through the dot.

In conclusion, the bolometric detectors show excellent electrical responsivity, higher than $10^9$ V/W, which is independent of frequency over an ultra-broadband spectrum, from sub-THz to ultraviolet. Even with larger (100 nm) dots, incident power as low as 30 pW at 1543 nm is easily detected by directly measuring the photovoltage $\Delta V_{DC}$. Although the responsivity decreases with increasing power, it is still at least three orders of magnitudes higher than other types of graphene bolometers [18,19] and it can be further optimized by using lock-in detection techniques. Future work will focus on pushing the performance limits of the device by using smaller quantum dots and optimizing the coupling of the incident radiation to the bolometer by designing optical cavities [27] and antennas [28].

ACKNOWLEDGEMENTS

This work was supported by the by the US Office of Naval Research (awards no. N000141310865 and N00014-16-1-2674 to Georgetown University and the University of Maryland as well as support to the US Naval Research Laboratory) and the NSF (ECCS 1610953). This research was performed while Kevin M Daniels held an NRC Research Associateship award at the US Naval Research Laboratory.

Appendix A: Methods

We used three light sources: a backward wave oscillator (Microtech) for the 2-mm wavelength radiation, a 1543-nm 81663A DFB Laser Source (Agilent) and a 365-nm LED UV source (U-Vix). The coupling efficiency for the 1543-nm wavelength source was estimated from the ratio of the absorbed power to the incident power. The incident power was estimated by multiplying the incident power density by the area of the graphene (20 $\mu m^2$, including the dot and the two triangular regions attached to it). The incident power density was measured with a calibrated detector 81524A InGaAs optical head with 8153A lightwave multimeter (HP).